\documentstyle[preprint,aps]{revtex}
\begin{document}
\title{Non-Ohmic Coulomb drag in the ballistic electron
transport regime}
\author{V. L. Gurevich and M. I. Muradov}
\address{\it Solid State Physics Department, A. F. Ioffe Institute,
194021 Saint Petersburg, Russia}
\maketitle
\begin{abstract}
We work out a theory of the Coulomb drag current created under
the ballistic transport regime in a one-dimensional nanowire by
a ballistic non-Ohmic current in a nearby parallel nanowire. As
in the Ohmic case, we predict sharp oscillation of the drag
current as a function of gate voltage or the chemical potential
of electrons.  We also study dependence of the drag current on
the voltage $V$ across the driving wire. For relatively large
values of $V$ the drag current is proportional to $V^2$.
\end{abstract}
The purpose of the present paper is to study the Coulomb drag
current in the course of ballistic (collisionless) electron
transport in a nanowire due to a ballistic driving non-Ohmic
current in an adjacent parallel nanowire. The possibility of the
Coulomb drag effect in the ballistic regime in quantum wires has
been demonstrated by Gurevich, Pevzner and
Fenton~\cite{GPF}\footnote{A number of references to early papers on
the Coulomb drag is given in~\cite{GPF}.} and has been experimentally
observed by Debray {\it et al.}~\cite{DVR}.  If two wires, 1 and 2, are
near one another and are parallel, the drag force due to the ballistic
current in wire 2 acts as a sort of permanent acceleration on the
electrons of wire 1 via the Coulomb interaction.

We assume that the largest dimension of the structure is smaller
than the electron mean free path in the problem (typically a few
$\mu$m). Such nanoscale systems are characterized by low
electron densities, which may be varied by means of the gate
voltage.

Following~\cite{GPF} we assume that the drag current in wire 1
is much smaller than the driving ballistic current in wire 2 and
calculate the drag current by iterating the Boltzmann equation for
wire 1. We have
\begin{equation}\label{df}
v{\partial F^{(1)}\over{\partial z}}=-I^{(12)}\{F^{(1)},F^{(2)}\},
\end{equation}
where $F^{(1,2)}$ are the electron distribution functions in
wires 1 and 2 respectively, $v=p/m$ is the electron velocity,
$p$ is the $x$-component of the electron quasimomentum, the $x$
axis is parallel to the wires.  The collision integral takes
into account only the interwire electron-electron scattering, so
that otherwise the electron motion in both wires is considered
as ballistic. Now,
\begin{equation}\label{ci} I^{(12)}\{F^{(1)},F^{(2)}\}=
\sum_{n^{\prime}p^{\prime}q}
W_{1pn,2p^{\prime}n^{\prime}}^{1p+qn,2p^{\prime}-qn^{\prime}}{\cal
P}
\end{equation}
where
\begin{equation}\label{3}
{\cal P}=\left[F^{(1)}_{np}F^{(2)}_{n^{\prime}p^{\prime}}
\left(1-F^{(1)}_{np+q}\right)\left(1-F^{(2)}_{n^{\prime}p^{\prime}-q}
\right)\right.-
\left.F^{(1)}_{np+q}F^{(2)}_{n^{\prime}p^{\prime}-q}
\left(1-F^{(1)}_{np}\right)\left(1-F^{(2)}_{n^{\prime}p^{\prime}}\right)
\right].
\end{equation}
As in Ref.~\cite{GPF}, we assume the wires to be different though
having the same lengths $L$ and consider the interaction processes when
electrons in the two nanowires after scattering remain within
the initial subbands
$\varepsilon^{(1)}_{np}=\varepsilon^{(1)}_n(0)+p^2/2m$ and
$\varepsilon^{(2)}_{n^{\prime}p}=\varepsilon^{(2)}_{n^{\prime}}(0)+p^2/2m$,
$n$ being the bands' number.

The first iteration of Eq.(\ref{df}) gives for the nonequilibrium part
of the distribution function $\Delta F_{np}^{(1)}$
\begin{equation}\label{df_1}
\Delta F^{(1)}_{np}=-\left(z\pm {L\over{2}}\right){1\over{v_{np}}}
I^{(12)}\{F^{(1)},F^{(2)}\}, \end{equation} for $p\,>\,0$ ($p\,<\,0$)
respectively. Using the particle conserving property of the scattering
integral
\begin{equation}\label{pcp}
\sum_{n}\int dpI^{(12)}\{F^{(1)},F^{(2)}\}=0
\end{equation}
we get for the drag current
\begin{equation}\label{current}
J=-2eL\sum_{n}\int_0^{\infty}
{dp\over2\pi\hbar}I^{(12)}\{F^{(1)},F^{(2)}\}.
\end{equation}

The scattering probability
$W_{1pn,2p^{\prime}n^{\prime}}^{1p+qn,2p^{\prime}-qn^{\prime}}$
in Eq.(\ref{ci}) includes a delta-function describing the energy
conservation for the electrons belonging to two different wires
\begin{eqnarray}\label{sp}
W_{1pn,2p^{\prime}n^{\prime}}^{1p+qn,2p^{\prime}-qn^{\prime}}
={2\pi\over{\hbar}}
\left|V_{1pn,2p^{\prime}n^{\prime}}^{1p+qn,2p^{\prime}-qn^{\prime}}\right|^2
\delta(\varepsilon^{(1)}_{np}+\varepsilon^{(2)}_{n^{\prime}p^{\prime}}-
\varepsilon^{(1)}_{np+q}-\varepsilon^{(2)}_{n^{\prime}p^{\prime}-q})
\end{eqnarray}
which following Ref.~\cite{GPF} can be brought into the form
\begin{equation}\label{delta_func}
\delta(\varepsilon^{(1)}_{np}+\varepsilon^{(2)}_{n^{\prime}p^{\prime}}-
\varepsilon^{(1)}_{np+q}-\varepsilon^{(2)}_{n^{\prime}p^{\prime}-q})=
{m\over{|p-p^{\prime}|}}\delta(q+p-p^{\prime}).
\end{equation}
This means that here we have backscattering processes and the
electrons swap their quasimomenta as a result of collision.

To calculate the drag current we do the first iteration of the
Boltzmann equation in the collision term. One can insert the
equilibrium distribution functions into the collision term, e.g.
$F^{(1)}_{np}=f(\varepsilon^{(1)}_{np}-\mu)$ for the first wire.
Here $f(\varepsilon-\mu)$ is the Fermi function.

We assume, in the spirit of the
Landauer-Buttiker-Imry~\cite{Lan,LB} approach, the driving
quantum wire to be connected to reservoirs which we call 'left`
$(l)$ and 'right` $(r)$, each of them being in independent
equilibrium described by shifted chemical potentials
$\mu^{(l)}=\mu-eV/2$ and $\mu^{(r)}=\mu+eV/2$. Here $\mu$ is the
average chemical potential while $\Delta\mu/e=V$ is the voltage
across wire 2 (we will assume that $eV>0$) and $e<0$ is the
electron charge.  Therefore, the electrons entering the wire
from the 'left` ('right`) and having quasimomenta $p^{\prime}>0$
($p^{\prime}<0$) are described by \begin{equation}\displaystyle
{F^{(2)}_{n^{\prime}p^{\prime}}=
f\left(\varepsilon^{(2)}_{n^{\prime}p^{\prime}}-\mu^{(l)}\right)\quad
p^{\prime}>0},\atop
\displaystyle{F^{(2)}_{n^{\prime}p^{\prime}}=
f\left(\varepsilon^{(2)}_{n^{\prime}p^{\prime}}-\mu^{(r)}\right)\quad
p^{\prime}<0} \label{a} \end{equation}
and we see that the collision
integral Eqs.~(\ref{ci}),
(\ref{3}) is identically zero if the initial quasimomentum
$p^{\prime}$ and the final quasimomentum $p^{\prime}-q$ of
electron are of the same sign. In other words we have here
backscattering processes; otherwise the equilibrium distribution
functions on the right-hand side of Eqs.~(\ref{ci}), (\ref{3})
would depend on the same chemical potential and the collision
term would vanish.

Due to Eq.(\ref{delta_func}) we will be interested only in the
values $p^{\prime}\,<\,0$ [because of the restriction
$p^{\prime}-q=p\,>0$ which follows from Eq.(\ref{current})]
and get the following product of distribution functions in the
collision term Eqs. (\ref{ci}) and (\ref{3})
\begin{equation}\label{ct}
{\cal P}=F^{(1)}_{np}F^{(2r)}_{n^{\prime}p^{\prime}}
\left(1-F^{(1)}_{np^{\prime}}\right)\left(1-F^{(2l)}_{n^{\prime}p}\right)
-F^{(1)}_{np^{\prime}}F^{(2l)}_{n^{\prime}p}
\left(1-F^{(1)}_{np}\right)\left(1-F^{(2r)}_{n^{\prime}p^{\prime}}\right),
\end{equation}
or
\begin{eqnarray}\label{ct_1}
{\cal
P}=f(\varepsilon^{(1)}_{np}-\mu)f(\varepsilon^{(2)}_{n^{\prime} p^{\prime}}
-\mu^{(r)})[1-f(\varepsilon^{(1)}_{np^{\prime}}-\mu) ]
[1-f(\varepsilon^{(2)}_{n^{\prime}p}-\mu^{(l)})]\\\nonumber
-f(\varepsilon^{(1)}_{np^{\prime}}-\mu)f(\varepsilon^{(2)}_{n^{\prime}p}
-\mu^{(l)})[1-f(\varepsilon^{(1)}_{np}-\mu)][1-
f(\varepsilon^{(2)}_{n^{\prime}p^{\prime}}-\mu^{(r)})].
\end{eqnarray}

This expression can be recast into the form
\begin{eqnarray}\label{ct_2}
{\cal
P}=2\sinh{\left({eV/{2k_{\rm B}T}}\right)}
\exp\{(\varepsilon^{(1)}_{np}-\mu)/k_{\rm B}T\}
\exp\{(\varepsilon^{(2)}_{n\prime p\prime}-\mu)/k_{\rm B}T\}\\
\times f(\varepsilon^{(1)}_{np}-\mu)
f(\varepsilon^{(2)}_{n^{\prime}p^{\prime}}-\mu-eV/2)\nonumber
f(\varepsilon^{(1)}_{np^{\prime}}-\mu)
f(\varepsilon^{(2)}_{n^{\prime}p}-\mu+eV/2)
\end{eqnarray}
and for the drag current we have
\begin{equation}
\label{drag_cur}
J=-2e\sinh{\left({eV\over{2k_{\rm B}T}}\right)}
{2\pi\over{\hbar}} {mL\over{2\pi\hbar}}
\left({2L\over{2\pi\hbar}}\right)^2
\left({2e^2\over {\kappa L}}\right)^2
\sum_{nn^{\prime}}\int_0^{\infty}dp\int_0^{\infty}dp^{\prime}
\displaystyle{g_{nn^{\prime}}(p+p^{\prime})\over{p+p^{\prime}}}
{\cal Q}
\end{equation}
where
\begin{equation}
\label{4}
{\cal Q}=\exp{\varepsilon^{(1)}_{np}-\mu\over k_{\rm B}T}
\exp{\varepsilon^{(2)}_{n\prime p\prime}-\mu\over k_{\rm B}T}
f(\varepsilon^{(1)}_{np}-\mu)f(\varepsilon^{(2)}_{n^{\prime}p^{\prime}}-
\mu-{eV\over2})
f(\varepsilon^{(1)}_{np^{\prime}}-\mu)
f(\varepsilon^{(2)}_{n^{\prime}p}-\mu+{eV\over2}),
\end{equation}
$\kappa$ is the dielectric susceptibility of the sample and
\begin{equation}\label{gnn}
g_{nn^{\prime}}(q)=\left[\int d{\bf r}_{\perp}\int d{\bf
r}^{\prime}_{\perp} |\phi_n({\bf r}_{\perp})|^2
K_0\left({|q|}|{\bf r}_{\perp}-{\bf r}^{\prime}_{\perp}|/\hbar\right)
|\phi_{n^{\prime}}({\bf r}^{\prime}_{\perp})|^2\right]^2 .
\end{equation}
Some estimates of this function are given in Ref.~\cite{GPF}.

According to the reasoning given in Ref.~\cite{GPF} all the terms of
the sum~(\ref{drag_cur}) where the differences
$\vert\varepsilon^{(1)}(0)-\varepsilon^{(2)}(0) \vert$ are much
bigger than both $k_{\rm B}T$ and $eV$ do not contribute to the
current $J$. Therefore we are left with the terms of the sum
where $\vert\varepsilon^{(1)}(0)-\varepsilon^{(2)}(0) \vert$
is smaller than or of the order of $k_{\rm B}T$ or $eV$. We will
assume for simplicity that there is only one such difference
(otherwise we would have gotten a sum of several terms of the same
structure).

As $\cal Q$ is a sharp function of $p$ and $p'$ one
can take out of the integral all the slowly varying
functions and get
\begin{equation}\label{part_case}
J=J_0\cdot{1\over2}\sinh\left({eV\over2k_{\rm B}T}\right)
\frac{\displaystyle{eV\over4k_{\rm B}T}-{\varepsilon_{nn'}\over 2k_{\rm B}T}}
{\sinh\left(\displaystyle{eV\over4k_{\rm B}T}-{\varepsilon_{nn'}\over
2k_{\rm B}T}\right)}
\cdot\frac{\displaystyle{eV\over4k_{\rm B}T}+{\varepsilon_{nn'}
\over 2k_{\rm B}T}}
{\sinh\left(\displaystyle{eV\over4k_{\rm B}T}+{\varepsilon_{nn'}\over
2k_{\rm B}T}\right)}
\end{equation}
where
\begin{equation}
J_0=-{8e^5m^3L(k_{\rm B}T)^2\over{\kappa^2\pi^2\hbar^4}}.
\displaystyle{g_{nn^{\prime}}(2p_n)\over{p_n^3}}
\end{equation}
Here we have introduced notation
$$\varepsilon_{nn^{\prime}}=
\varepsilon^{(1)}_{n}(0)-\varepsilon^{(2)}_{n^{\prime}}(0),\quad
mv_n=p_n=\sqrt{2m[\mu-\varepsilon^{(1)}_n(0)]}.$$

Let us give an order-of-magnitude estimate of the drag current
$J$ for a realistic situation. We assume $T=$1K, $\mu$=14 meV,
the widths of the wires are 25 nm, the distance between thr
central lines of the wires is 50 nm, $\kappa$=13,
$m=6.7\cdot10^{-29}$. Then
\begin{equation}
J_0\approx10^{-10} \rm A .
\label{6}
\end{equation}

For $eV\ll k_{\rm B}T$ one gets from Eq. (\ref{part_case}) the result of
Ref.~\cite{GPF}. Let us consider in detail the opposite case
$eV\gg k_{\rm B}T$.
In this case one gets a nonvanishing result for
Eq.(\ref{part_case}) only if
$|\varepsilon_{nn\prime}|\,<\,eV/2$
and one obtains the following equation for the drag current
\begin{equation}
J={\cal B}
\left[\left({eV\over2}\right)^2-
\left({\varepsilon_{nn'}}\right)^2\right],\quad
{\cal B}=-{2e^5m^3L\over{\kappa^2\pi^2\hbar^4}}
\cdot\displaystyle{g_{nn^{\prime}}(2p_n)\over{p_n^3}}.
\label{2}
\end{equation}
The situation is illustrated in Fig. 1. The straight lines
correspond to the positions of the chemical potentials
$\mu^{(r)}$ and $\mu^{(l)}$ while the dashed line corresponds to
the average value $\mu$. Parabolas (1) and (2) represent the
dispersion law of electrons in wires (1) and (2) respectively.
The full circles correspond to the initial states of colliding
electrons.

{\it Before the collision} states 1a and 2a are occupied. The circle
representing state 1a is below the dashed line, i.e. below the
Fermi level $\mu$. The circle 2a represents a state with $p>0$
which is also occupied as the corresponding energy is below
$\mu^{(r)}$.

{\it After the collision} state 1b is occupied. It is
represented by a circle above the dashed line which means that
it has been free before the collision. In wire 2 state 2b with
$p<0$ is also occupied. It is above $\mu^{(l)}$, i.e. it has
been free before the transition.

The width of the stripe between the two straight lines is $eV$.
The drag current should be proportional to the number of the
occupied initial states as well as to the number of free final
states. As a result we have for $eV\gg|\varepsilon_{nn'}|$,
$\quad J\propto V^2$ --- see Fig. 2.

In summary, we have developed a theory of Coulomb drag current
in a quantum wire brought about by a non-Ohmic current in a
nearby parallel nanowire. A ballistic transport in both
nanowires is assumed. The drag current $J$ as a function of the
gate voltage comprises a system of spikes; the position of each
spike is determined by a coincidence of a pair of levels of
transverse quantization, $\varepsilon_{n}(0)$ and
$\varepsilon_{n'}(0)$ in both wires.  For $eV\gg k_{\rm B}T$,
$J$ is a parabolic function of the driving voltage $V$. The effect
may play an important role in the investigation of the interwire
Coulomb scattering as well as 1D band structure of the wires. 

The authors are grateful to P. Debray for sending them a preprint of
paper~\cite{DVR} prior to publication. The authors are pleased to
acknowledge the support for this work by the Russian National Fund of
Fundamental Research (Grant No~97-02-18286-a).

\bigskip
\centerline{FIGURE CAPTIONS}

\bigskip
Fig. 1. Schematic representation of simultaneous transitions due
to the interaction between electrons of the two wires for $eV\gg
k_{\rm B}T$. Circles $\circ$ and $\bullet$ represent the initially
unoccupied and occupied states respectively.

\bigskip
Fig. 2. Dependence of the drag current on the driving voltage for
$T=$1 K. The values of $|\varepsilon_{nn'}|$ go up from left to right.
\end{document}